\documentclass[11pt]{article}
\usepackage{amsmath,enumerate,amsfonts,color,url,amssymb,ntheorem,cite}%,showlabels}

{\catcode `\@=11 \global\let\AddToReset=\@addtoreset}
\AddToReset{equation}{section}

\AddToReset{figure}{section}

\newcounter{mnotecount}[section]
\renewcommand{\themnotecount}{\thesection.\arabic{mnotecount}}

\DeclareFontFamily{OT1}{rsfs}{}
\DeclareFontShape{OT1}{rsfs}{m}{n}{ <-7> rsfs5 <7-10> rsfs7 <10-> rsfs10}{}
\DeclareMathAlphabet{\mycal}{OT1}{rsfs}{m}{n}

\definecolor{MDB}{rgb}{0,0.08,0.45}
\definecolor{MyDarkBlue}{rgb}{0,0.08,0.45}

\definecolor{MLM}{cmyk}{0.1,0.8,0,0.1}
\definecolor{MyLightMagenta}{cmyk}{0.1,0.8,0,0.1}

\definecolor{HP}{rgb}{1,0.09,0.58}

\newcommand\mcMloc{{\mcM_{\rm loc}}}
\newcommand\rhoS{{\rho_S}}
\newcommand\zS{{z_S}}
\newcommand\zetaS{{\zeta_S}}
\newcommand\ringX{{\mathring{X}}}

\newcommand{\jlcax}[1]{}
\newcommand{\eean}{\nonumber\end{eqnarray}}

\newcommand{\kk}[1]{}%{\mnote{{\bf If we consider the KK case:} #1}}

\newcommand{\mcH}{{\mycal H}}

\newcommand{\beq}{\begin{equation}}

\newcommand{\FS}       %{F_1} %
                  {F}
\newcommand{\HS} %{F_2}
       {H_{\mbox{\scriptsize volume}}}

{\ptc{this should be removed in the oberwolfach version}}%

\newcommand{\mcA}{{\mycal A}}%
\newcommand{\eeal}[1]{\label{#1}\end{eqnarray}}
\newcommand{\bed}{\begin{deqarr}}
\newcommand{\eed}{\end{deqarr}}
\newcommand{\bedl}[1]{\begin{deqarr}\label{#1}}
\newcommand{\eedl}[2]{\arrlabel{#1}\label{#2}\end{deqarr}}

\newcommand{\bel}[1]{\begin{equation}\label{#1}}
\newcommand{\bea}{\begin{eqnarray}}
\newcommand{\bean}{\begin{eqnarray}\nonumber}
\newcommand{\beal}[1]{\begin{eqnarray}\label{#1}}
\newcommand{\eea}{\end{eqnarray}}

\newcommand{\be}{\begin{equation}}
\newcommand{\eeq}{\end{equation}}
\newcommand{\ee}{\end{equation}}
\newcommand{\beqa}{\begin{eqnarray}}
\newcommand{\eeqa}{\end{eqnarray}}
\newcommand{\beqan}{\begin{eqnarray*}}
\newcommand{\eeqan}{\end{eqnarray*}}
\newcommand{\ba}{\begin{array}}
\newcommand{\ea}{\end{array}}

\newcommand{\hyp}{\mycal S}
\newcommand{\HH}{{\cal H}}

\newcommand{\mcM}{{\mycal M}}
\newcommand{\mcD}{{\mycal D}}

\newcommand{\scri}{{\mycal I}}%
\newcommand{\scrip}{\scri^{+}}%

\newcommand{\mnote}[1]%{}
{\protect{\stepcounter{mnotecount}}$^{\mbox{\footnotesize
$%\!\!\!\!\!\!\,
\bullet$\themnotecount}}$ \marginpar{%\color{red}%
\raggedright\tiny\em
$\!\!\!\!\!\!\,\bullet$\themnotecount: #1} }

\newcommand{\warn}[1]%{}%{}
{\protect{\stepcounter{mnotecount}}$^{\mbox{\footnotesize
$%\!\!\!\!\!\!\,
\bullet$\themnotecount}}$ \marginpar{%\color{red}%
\raggedright\tiny\em
$\!\!\!\!\!\!\,\bullet$\themnotecount: {\bf Warning:} #1} }

\newcommand{\R}{\mathbb R}

\newcommand{\eq}[1]{(\ref{#1})}

\newcommand{\Mext}{M_\ext}
\newcommand{\ext}{\mathrm{ext}}

\newcommand{\ptc}[1]{\mnote{{\bf ptc:}#1}}

\newcommand{\mcS}{{\mycal S}}
\newcommand{\mcL}{{\mycal L}}

\newcommand{\beqar}{\begin{deqarr}}
\newcommand{\eeqar}{\end{deqarr}}

\newcommand{\beaa}{\begin{eqnarray*}}
\newcommand{\eeaa}{\end{eqnarray*}}

\newcommand{\tr}{\mbox{tr}}

\def\RR{{\mathbb R}}
\def\HH{{\mathbb H}}

\def\FS{{\mathfrak S}}

\def\mcHdeg{{\mcH_{\rm deg}}}

\newtheorem{Theorem} {\sc  Theorem\rm} [section]

\newtheorem{proposition} [Theorem] {\sc  Proposition\rm}
\newtheorem{theorem}[Theorem]{\sc  Theorem\rm}

\theorembodyfont{\upshape}

\newcounter{marnote}

\begin{document}

\title{Existence of singularities in two-Kerr black holes\thanks{Preprint UWThPh-2011-24}}
\author{Piotr T. Chru\'{s}ciel\thanks{Gravitational Physics, University of Vienna},
Micha\l\  Eckstein\thanks{Faculty of Mathematics and Computer Science, Jagellonian University and Copernicus Center},
Luc Nguyen\thanks{Department of Mathematics, Princeton University} and
Sebastian J. Szybka\thanks{Astronomical Observatory,  Jagellonian University and Copernicus Center}}
\date{}
\maketitle

\begin{abstract}
We show that the angular momentum -- area inequality $8\pi |J|\le A$ for weakly stable minimal surfaces would apply to $I^+$-regular many-Kerr solutions, if any existed.  Hence we remove the undesirable hypothesis in the Hennig-Neugebauer proof of non-existence of well behaved two-component solutions.
\end{abstract}

%\tableofcontents

%----------------------------------------------------------------------------%
%\section{Introduction}

%\renewcommand{\ptcr}[1]{}

\section{Introduction}

An intriguing open problem in the theory of stationary axisymmetric vacuum black holes is that of existence of regular solutions with disconnected components. Key progress on this was made recently by Hennig and Neugebauer~\cite{HennigNeugebauer3} (see also~\cite{Neugebauer:2003qe,Varzugin1}) concerning two-components solutions:

\begin{enumerate}
 \item According to~\cite{{Neugebauer:2003qe,HennigNeugebauer,HennigNeugebauer3,Varzugin1}}, if a sufficiently regular multi-component solution exists, it must belong to the multi-Kerr family.
      \item There are \emph{no}  double-Kerr solutions satisfying a \emph{``sub-extremality"} condition~\cite{HennigNeugebauer,HennigNeugebauer2,HennigNeugebauer3}.
\end{enumerate}

The sub-extremality condition of~\cite{HennigNeugebauer,HennigNeugebauer2,HennigNeugebauer3} appears as an undesirable restriction on the class of space-times considered. It is the purpose of this work to tie instead the analysis of Hennig and Neugebauer to stability properties of Killing horizons,
and to point out that the stability condition is necessarily satisfied by, say, $I^+$-regular black hole space-times.
More precisely, in Section~\ref{S28VII11.1} below we establish the following result:

\begin{theorem}
 \label{T28VII11.1}
{$I^+$-regular two-Kerr solutions do not exist}.
\end{theorem}

Some more comments on the issues arising might be in order. The key use of the sub-extremality condition in~\cite{HennigNeugebauer,HennigNeugebauer2,HennigNeugebauer3} is a result from \cite{HennigAnsorgCederbaum} which states that the horizon area $A$ and the horizon angular momentum $J$ satisfy $A > 8\pi|J|$. Interestingly, a recent result of~\cite{DainReiris} (see also~\cite{JRD}) asserts that an alternative condition on the horizon implies a weaker area-angular momentum inequality $A \geq 8\pi|J|$. This naturally begs for clarifications
\begin{enumerate}[(a)]
  \item whether the sub-extremality condition in~\cite{HennigNeugebauer,HennigNeugebauer2,HennigNeugebauer3} is related to the conditions in~\cite{DainReiris},
  \item and whether some natural global properties of space-times considered imply that horizons   satisfy the condition.
\end{enumerate}
These are the main issues addressed by this work. In fact, our proof may be viewed as a verification that~\cite{DainReiris} applies to the problem at hand, i.e. the answer to (b) is positive for $I^+$-regular double-Kerr black holes. Regarding (a),  while the proofs of the area-angular momentum inequality  in~\cite{HennigAnsorgCederbaum} and in~\cite{DainReiris} are different, they share the same intermediate step: the inequality  (15) in~\cite{HennigAnsorgCederbaum} is the strict version of the inequality (34) in \cite{DainReiris}. In any event,  we give in Appendix~\ref{A15VII11.1} a self-contained derivation of the inequality as needed for our purposes, using an argument closely related to, but not identical with, that in~\cite{HennigAnsorgCederbaum}.

%----------------------------------------------------------------------------%
\section{The proof}
 \label{S28VII11.1}

Arguing by contradiction, consider (regular) metrics of the form
\bel{10VII11.1}
 g= f^{-1}( h (d\rho^2 + dz^2 ) + \rho^2 d\varphi^2 ) - f (dt+ a d\varphi)^2
%  \;,
\ee
on
\bel{28VII11.1}
 \mbox{$\mcM:=\{t,z\in \R$, $\varphi\in [0,2\pi]
 $, $\rho\in [0,\infty)\}$,
}
\ee
and where $f$ takes the explicit form considered in~\cite{HennigNeugebauer3}. The potential $a$ can be obtained from~\cite[Eq.~(30)]{HennigNeugebauer3} (the auxiliary function $
\chi$ needed for this can be found in Eq.~(92) there, while $\psi$ can be obtained from $\chi$ using Eq.~(26) there).  The function  $h$ can be calculated using \cite[Eq. (15)]{Kramer}.%
\footnote{We are grateful to J.~Hennig for pointing out the relevant equations, and for making his {\sc Maple} files available to us.}
This results in a manifestly stationary and axisymmetric metric: $\partial_t g_{\mu\nu}=0=\partial_\varphi g_{\mu\nu}$.

We start by noting that double-Kerr solutions with vanishing surface gravity of both components (in the notation below, this corresponds to $K_1=K_2$ and $K_3=K_4$, whence $S=\emptyset$) have been shown to be singular in~\cite{HennigNeugebauer2} without supplementary hypotheses, and do not require further considerations. It remains to consider the case where both components have non-vanishing surface gravity or exactly one of the components has non-vanishing surface gravity.

 We assume that the metric \eq{10VII11.1} describes the \emph{domain of outer communications}
of a well behaved vacuum space-time with an event horizon which has two components. More precisely, we \emph{assume} that there exists a choice of the parameters occurring in the metric functions such that the following conditions hold. First, we assume that
\begin{enumerate}[1.]
\item The metric functions $f$, $af$, $a^2f-f^{-1}\rho^2$ and $hf^{-1}$ are smooth for $\rho>0$.

\item The Killing vector $\partial_\varphi$ is spacelike wherever non-vanishing. In particular
$$
 g(\partial_\varphi,\partial_\varphi) = f^{-1} \rho^2 - f a^2 > 0 \ \mbox{for} \ \rho>0
 \;.
$$
\end{enumerate}
In case that both components of the horizon are non-degenerate, i.e. they have non-vanishing surface gravity, we supplement the above with an assumption that there exists $K_1 > K_2 > K_3 > K_4$ such that
\begin{enumerate}[1.]
\addtocounter{enumi}{2}
\item On the boundary
$$\mcA:=\{\rho=0\}
	\;,
$$
the intervals defined by $z\not\in[K_1,K_2] \cup [K_3,K_4]$ can be mapped, by a suitable coordinate transformation, to a smooth axis of rotation for the Killing vector $\partial_\varphi$.
\item There exists a coordinate system in which the manifold $\hyp_0:=\{t=0\}$ equipped with the metric %
\bel{10VII11.2}
  f^{-1}( h (d\rho^2 + dz^2 ) + \rho^2 d\varphi^2 ) - f (a d\varphi)^2
%  \;,
\ee
is a smooth Riemannian manifold whose boundary is located exactly at the intervals $z \in[K_1,K_2]$ and $z \in [K_3,K_4]$ such that each of these intervals corresponds to a smooth sphere.
\item Finally, the space-time manifold $\mcM$ defined in \eq{28VII11.1} can be extended so that
   the boundary spheres
\bel{14VII11}
 \mbox{$S_1:=\{t=0,z \in[K_1,K_2]\}$ and $S_2:=\{t=0,z \in [K_3,K_4]\}$}
 \ee
 are bifurcation surfaces of  bifurcate Killing horizons.
\end{enumerate}
If one of the component of horizon is degenerate (i.e. it has vanishing surface gravity) and one is non-degenerate, we assume that there exist $K_1 > K_2 > K_3 = K_4$ such that
\begin{enumerate}[1'.]
\addtocounter{enumi}{2}
\item
On the boundary $\mcA:=\{\rho=0\}$, the intervals defined by $z\not\in[K_1,K_2] \cup \{K_3\}$ can be mapped, by a suitable coordinate transformation, to a smooth axis of rotation for the Killing vector $\partial_\varphi$.
\item There exists a coordinate system in which the manifold $\hyp_0:=\{t=0\}$ equipped with the metric %
\bel{10VII11.2asdf}
  f^{-1}( h (d\rho^2 + dz^2 ) + \rho^2 d\varphi^2 ) - f (a d\varphi)^2
%  \;,
\ee
is a smooth Riemannian manifold which has a boundary located at the interval $z \in[K_1,K_2]$ corresponding to a smooth sphere and an asymptotically cylindrical end located at $z = K_3$.
\item Finally, the space-time manifold $\mcM$ defined in \eq{28VII11.1} can be extended so that
   the boundary sphere
\bel{14VII11asdf}
 \mbox{$S_1:=\{t=0,z \in[K_1,K_2]\}$}
 \ee
is a bifurcation surface of a bifurcate Killing horizon while $z = K_3$ corresponds to a smooth degenerate Killing horizon.
\end{enumerate}
In point 4', we adopt the following definition for a cylindrical end: An initial data set $([0,\infty)\times N,g,K)$ is called a \emph{cylindrical end} if $N$ is a compact manifold, if the metric $g$ approaches an $x$--independent Riemannian metric on
$[0,\infty)\times N$  as the variable $x$ running along the $ [0,\infty)$--factor tends to infinity,   and if the extrinsic curvature tensor $K$ approaches an $x$-independent symmetric tensor field as $x$ tends to infinity.

The seemingly \emph{ad hoc} requirements set forth above can be derived from the hypothesis of \emph{$I^+$-regularity} of the two-component black-hole axisymmetric configuration under consideration, see \cite{ChCo,ChNguyenMass,ChNguyen} and references therein.

Recall that, in vacuum, the \emph{stability operator} for a  marginally outer trapped surface (MOTS) is defined as
\bel{16VII11.21}
 -\Delta_S \phi + 2 K(\nu,\mcD \phi) + \big(\mbox{\rm div}_S K(\nu, \cdot) - \frac 12 |\chi |^2 - |K(\nu, \cdot)|^2 + \frac 12 R_S \big)\phi
 \;,
\ee
where $\phi$ is a smooth function on $S$, $\Delta_S$ is the Laplace operator on $S$, $\nu$ is a field of unit normals, $\mcD$ is the Levi-Civita derivative of the metric on $S$, $\mbox{\rm div}_S$ is the divergence on $S$, and $R_S$ is the scalar curvature of the metric induced on $S$. Finally, if $W$ is the second fundamental form of $S$, then $\chi$ is defined as $W+K_S$, where $K_S$ is the pull-back of $K$ to $S$. Following~\cite{AndM2,AMS2}, a MOTS will be called \emph{stable} if the smallest real part of all eigenvalues of the stability operator is non-negative.

It follows from Appendix~\ref{A14VII11.1} that, on $S$, $K(\nu, \cdot)$ is proportional to $d\varphi$, and hence $K(\nu, \mcD \phi)$  vanishes for {axisymmetric}  functions $\phi$. The divergence $\mbox{\rm div}_S K(\nu, \cdot)$ vanishes for similar reasons. Since all the metric functions are smooth functions of $\rho^2$ away from the axis of rotation, one easily finds that $W$ vanishes. So, on $\varphi$-independent functions, the stability operator reduces to
\bel{16VII11.21asdf}
 -\Delta_S \phi   +\frac 12  \big(  R_S  -  |K|^2  \big)\phi
 \;,
\ee
which, after using the vacuum constraint equation on the maximal hypersurface $\hyp_0$, is the familiar stability operator for minimal surfaces \emph{within the class of functions that are invariant under rotations}.

We note the following
theorem, pointed out to us by  M.~Eichmair (private communication), which follows from the existence theory for marginally outer trapped
surfaces\cite{AEM} with an additional argument to
accommodate the cylindrical ends:

\begin{Theorem}
 \label{T13VII1.1}
Consider a smooth initial data set $(M,g,K)$, where $M$ is the union of a compact set with several asymptotic ends in which the metric is either asymptotically flat or asymptotically cylindrical, with at least one asymptotically flat end. If $M$ has a non-empty boundary, assume that $\partial M$ is weakly inner-trapped, where ``inner" refers to variations pointing towards $M$.
If $|K|$ tends to zero as one recedes to infinity along the cylindrical ends (if any), then  there exists an outermost smooth compact MOTS in $M$ which is \underline{stable}.
\end{Theorem}

Indeed, this theorem is one of the main results in \cite{AndM2,Eichmair}
when there are only asymptotically flat ends. In the setting above one can deform each cylindrical end  to an asymptotically flat end  for $x\ge x_0$, and apply the asymptotically flat version of the theorem to the deformed data set. When $x_0$ has been chosen large enough one shows that the resulting stable smooth compact MOTS  is contained in the region $x\le x_0$, and hence provides an stable MOTS within the original undeformed data set.
This argument uses the asymptotic vanishing of $|K|$ in the cylindrical ends. Unfortunately this last condition is too strong to be useful for a direct argument which only uses the $\{t=0\}$ slice. An extension of Theorem~\ref{T13VII1.1} to cylindrical ends with tensors $K$ as described in Appendix~\ref{A14VII11.1} would immediately extend our analysis of the $\{t=0\}$ slice to configurations with one horizon degenerate and one not. (In fact, one would only need an extension of this theorem to axially symmetric MOTS, in which case various terms in the relevant equations vanish, possibly simplifying the analysis.) In any event, the version in \cite{AndM2,Eichmair} suffices in our subsequent analysis.

Going back to our problem, let us first treat the case where the horizon has two non-degenerate components: we assume that the boundaries  $S_1$ and $S_2$ of $\mcS_0 = \{t = 0\}$ defined in \eq{14VII11} are non-empty. Now, it is standard that each $S_a$ has vanishing null expansions, in particular it is weakly outer-trapped; by the latter we mean that the future outer null expansion of $S_a$ is non-negative.
We want to show that each $S_a$ is \emph{stable} in the sense of~\cite{AndM2}.
For this, suppose first that $S:=S_1\cup S_2$ is not \emph{outermost} as a MOTS. By  Theorem~\ref{T13VII1.1} (without cylindrical ends) $\hyp_0$ contains a smooth compact MOTS, say $S'$, which must be distinct from $S$ since $S'$ is  stable and $S$ is not. In particular $S'$   contains a smooth compact component, say $S'_1$, which is distinct from both $S_a$'s. By~\cite[Theorem~6.1]{CGS}  the MOTS $S_1'$ cannot be seen from $\scrip$, hence $\hyp_0$ is not contained in the domain of outer communications, which is a contradiction. Thus $S$ is a stable MOTS.

On the other hand, we have just seen that the MOTS-stability operator of $S$ coincides with its minimal-surface stability operator.
  We deduce that $S$ is a minimal surface within $\hyp_0$ which is weakly stable, in the sense of minimal surfaces contained in $\hyp_0$,  with respect to variations invariant under rotations around an axis of symmetry.

   In other words, it holds
\[
\int_{S} \Big\{|\nabla_S \phi|^2 + \frac{1}{2} (R_S - R_{\mcS_0})\phi^2\Big\}\,dv_S \geq 0 \text{ for all } \phi \in C^\infty(S), \partial_\varphi \phi = 0
\]
(compare Equation (2.12) in~\cite{GallowaySchoen}).
This allows us to invoke~\cite{DainReiris}  (see also Appendix~\ref{A15VII11.1} below) to conclude that each component of $S$ satisfies
\bel{26VII11.1}
 A_a \ge 8\pi |J_a|
 \;,
\ee
where $A_a$  is the area of $S_a$ and $J_a$ its Komar angular momentum. But this inequality has been shown to be violated by at least one of the components%
\footnote{Actually Hennig and Neugebauer draw contradictions from $A _a > 8\pi |J_a|$, but  the inequality \eq{26VII11.1} suffices.} of $S$ by Hennig and Neugebauer~\cite{HennigNeugebauer}, which gives a contradiction, and proves Theorem~\ref{T28VII11.1}.

Consider, finally, a two-Kerr metric with exactly one non-degenerate component, and assume that the space-time is $I^+$-regular. As already mentioned, Theorem \ref{T13VII1.1} does not apply directly to the hypersurface $\{t = 0\}$ which contains a cylindrical end. Instead we argue as follows. Let $\hyp$ be any hypersurface in the d.o.c. with compact boundary on a degenerate component of $\partial J^-(\Mext) \cap I^+ (\Mext)$ (for notation, see~\cite{ChCo}), and which coincides with the surface $\{t=0\}$ near the non-degenerate component of the event horizon. One possible method to obtain such hypersurface is to start with a spacelike, acausal hypersurface given by the $I^+$-regularity assumption and deform it near the birfurcate horizon to $\{t = 0\}$ using the construction in \cite{RaczWald2}.  We also detail a different construction for one such hypersurface in Appendix \ref{AppSurfConstr}, which is more elementary in nature. Arguing exactly as before, both components of $\partial \hyp$ are stable in the sense of MOTs. It follows as above that the non-degenerate component of the event horizon satisfies \eq{26VII11.1}. Now, for such configurations Hennig and Neugebauer show that the inequality $
 A _1 > 8\pi |J_1|$ implies that the total mass $m$ of the metric is strictly negative. One can check
(J.~Hennig, private communication)
that under \eq{26VII11.1} the conclusion still holds, which is incompatible with the positive energy theorem for black holes~\cite{Herzlich:bh}. Hence no such configurations are possible, and the theorem is established.

\appendix
 \section{The extrinsic curvature tensor}
  \label{A14VII11.1}

For a metric of the form \eq{10VII11.1} the field $N$ of future-pointing unit normals to $\hyp_0$ reads
$$
 N = \sqrt{\frac{\rho^2 - a^2 f^2}{f \rho^2}} \bigg(\partial_t + \frac{a f^2}{\rho^2 -a^2 f^2}\partial_\varphi\bigg)
 \;.
$$
Since $N$ has only $t$ and $\varphi$ components, the hypersurface $\hyp_0$ is maximal:
$$
 \tr K = \nabla_\mu N^\mu = \frac 1 {\sqrt{|\det g|}}\big( \partial_t( \sqrt{|\det g|}N^t)
 +\partial_\varphi( \sqrt{|\det g|}N^\varphi)\big)
 = 0
 \;.
$$
It further follows that the extrinsic curvature tensor $K_{ij}= \frac 12 \mcL_N g_{ij}$ equals
\beaa
 K_{ij} dx^i dx^j  &=& \frac 12 \big( N^\mu \partial_\mu g_{ij}
  + \partial_i N^\mu g_{\mu j}
  + \partial_j N^\mu g_{\mu i} \big) dx^i dx^j
\\
 & = &
 \partial_z N^\mu g_{\mu \varphi} dz d\varphi
  + \partial_\rho N^\mu g_{\mu \varphi } d\rho d\varphi
 \;.
\eeaa
This leads to the following non-vanishing components of $K_{ij}$
\beal{11VII11.5}
 K_{z\varphi}=K_{\varphi z} &= & \frac{f(\rho ^2   +a^2
     f^2)\partial_z a+2 \rho ^2 a
   \partial_z f }{2\rho
   \sqrt{f \rho ^2-a^2
   f^3}}
   \;,
\\
 K_{\rho\varphi}=
 K_{\varphi\rho} &= &
\frac{f(\rho ^2  +a^2
    f ^2)\partial_\rho a+2 \rho  a
   \left(\rho  \partial_\rho f -f\right)}{2\rho
   \sqrt{f\rho ^2-a^2
   f^3}}
\;.
\eeal{11VII11.6}

One can invoke the construction of the $(\rho,z)$ coordinates in~\cite{ChUone} to show that, for a smooth hypersurface $\hyp_0$, the limits $\lim_{\rho\to 0} f$ and  $\lim_{\rho\to 0} h$ are smooth strictly negative functions of $z$   on $(K_1,K_2)$ and $(K_3,K_4)$. Similarly  $\lim_{\rho\to 0} a$ exists and is constant on each of $(K_1,K_2)$ and $(K_3,K_4)$. (Its value on these two intervals might be different.)

Using that, away from the $K_i$'s, all the functions appearing in the metric are smooth functions of $\rho^2$, that $a$ is constant on the horizon, and that $f$ is strictly negative near the horizon away from the axis,
at the bifurcation surface $\rho=0$ we obtain
\beal{11VII11.5a}
 K_{z\varphi}=K_{\varphi z} &= & 0
   \;,
\\
 K_{\rho\varphi}=
 K_{\varphi\rho} &= & - \lim_{\rho\to0} \left(
\frac{ a^2
    f ^2 \partial_\rho a }{2\rho
   \sqrt{ -a^2
   f }}-
\frac{a}{  \sqrt{-
   a^2f }}\right)
\;.
\eeal{11VII11.6a}

\section{The area inequality}
 \label{A15VII11.1}

Consider an axially symmetric, stationary smooth vacuum metric
\begin{equation}
g = -\frac{\rho^2}{X}\,dt^2 + X\,(d\varphi - \omega\,dt)^2 + \frac{e^{2\mu}}{X}(d\rho^2 + dz^2)
	\label{Eq.MetricgLoc}
\end{equation}
defined on $\mcMloc = (U \setminus \Gamma) \times \RR$ where $\Gamma$ is a finite line segment centered at the origin and of length $2m>0$ lying on the $z$-axis of $\RR^3$, $U$ is an open neighborhood of $\Gamma$ in $\RR^3$, $(\rho,\varphi,z)$  {are} the cylindrical coordinates of the $U \setminus \Gamma$-factor, $t$ is the time function on the $\RR$-factor and all relevant functions are independent of both $t$ and $\varphi$. The subset $\mcH = \Gamma \times \RR$ of $\{\rho = 0\}$ is the ``event horizon" of $\mcMloc$ and its complement $\mcA$ (relative to $\{\rho = 0\}$) is the axis of rotation of $\mcMloc$.

The smoothness of the metric $g$ is assumed to be up to its boundary in the following sense: There exists a coordinate transformation $\Psi$ from a neighborhood $U$ of $\Gamma$ into a subset $V$ of $\RR^3$ such that the line segment $\Gamma$ is unwrapped to a smooth sphere and the metric $g$ relative to the new coordinate system is smooth up to that sphere. Note that, as
shown in detail in~\cite{ChNguyenMass},%
\footnote{When working on the current paper, we (PTC and LN) have realised that these coordinates have been already used in~\cite{Bardeen}; there the reader is directed to unpublished work of Carter.}
the transformation $\Psi$ can be taken to be the Joukovsky transformation
\[
\zeta = \rho + z \mapsto \rhoS + i\,\zS = \zetaS \ \text{ such that  } \zeta = \zetaS + \frac{m^2}{4\zetaS}.
\]
Recall that the midpoint of $\Gamma$ is the origin and the length of $\Gamma$ is $2m$. We also use the coordinate functions $r$ and $\theta$ defined by $\rhoS = r\sin\theta$ and $\zS = r\cos\theta$. The image of $\Gamma$ under this transformation is $\Sigma = \{r = \frac{m}{2}\}$. With respect to $(t,r,\varphi, \theta)$, the metric $g$ takes the form
\bel{14VII11.11}
g = -\frac{\rho^2}{X}\,dt^2 + X\,(d\varphi - \omega\,dt)^2 + \frac{e^{2\mu}}{X}\underbrace{\frac{\big(r^2\,\cos 2\theta -  \frac{m^2}{4}\big)^2 + r^4\,\sin^2 2\theta}{r^4}}_{=: \Xi}(dr^2 + r^2\,d\theta^2)
\ee
where, by a standard abuse of notation, we use the same symbol for $X(r,\theta)$ and $ X(\rho,z)$, etc.

Away from $\{\rho = 0\}$ the Einstein vacuum equations are equivalent to (see, e.g., \cite[Proposition~4.1]{Heusler:book})
\begin{align}
&(\rho\,X_{,\rho})_{,\rho} + (\rho\,X_{,z})_{,z} = \rho\,\frac{X_{,\rho}^2 + X_{,z}^2 - Y_{,\rho}^2 - Y_{,z}^2}{X}
	\;,\label{X}\\
&(\rho\,Y_{,\rho})_{,\rho} + (\rho\,Y_{,z})_{,z} = 2\rho\,\frac{X_{,\rho}\,Y_{,\rho} + X_{,z}\,Y_{,z}}{X},\label{Y}\\
&\omega_{,\rho} = - \frac{\rho}{X^2}\,Y_{,z}
	\;,\label{Omega,rho}\\
&\omega_{,z} = \frac{\rho}{X^2}\,Y_{,\rho}
	\;,\label{Omega,z}\\
&\mu_{,\rho} = \frac{\rho}{4X^2}(X_{,\rho}^2 - X_{,z}^2 + Y_{,\rho}^2 - Y_{,z}^2)
	\;,\label{lambda,rho}\\
&\mu_{,z} = \frac{\rho}{2X^2}(X_{,\rho}\,X_{,z} + Y_{,\rho}\,Y_{,z})
	\;,\label{lambda,z}
\end{align}
where $Y$ is an important auxiliary function, usually thought of as the imaginary part of the complex Ernst potential $X+iY$ associated with the rotational Killing field $\partial_\varphi$. (We emphasise that here we use the Ernst potential related to the rotational Killing vector and not the asymptotically timelike one.) Equivalently, we have
\begin{align}
&\frac{1}{r^2 - \frac{m^2}{4}}\Big(\big(r^2 - \frac{m^2}{4}\big)\,X_{,r}\Big)_{,r} + \frac{1}{r^2\,\sin\theta}\Big(\sin\theta\,X_{,\theta}\Big)_{,\theta}\nonumber\\
	&\qquad\qquad = \frac{X_{,r}^2 + r^{-2}X_{,\theta}^2 - Y_{,r}^2 - r^{-2}Y_{,\theta}^2}{X}
	\;,\label{Xrt}\\
&\frac{1}{r^2 - \frac{m^2}{4}}\Big(\big(r^2 - \frac{m^2}{4}\big)\,Y_{,r}\Big)_{,r} + \frac{1}{r^2\,\sin\theta}\Big(\sin\theta\,Y_{,\theta}\Big)_{,\theta}\nonumber\\
	&\qquad\qquad = 2\,\frac{X_{,r}\,Y_{,r} + r^{-2}\,X_{,\theta}\,Y_{,\theta}}{X}
	\;,\label{Yrt}\\
&\omega_{,r} = - \frac{\rho}{r\,X^2}\,Y_{,\theta}
	\;,\label{Omega,r}\\
&\omega_{,\theta} = \frac{\rho\,r}{X^2}\,Y_{,r}
	\;,\label{Omega,theta}\\
&\mu_{,r}
	= \frac{\rho}{4X^2\,\Xi}\Big[\frac{\sin\theta(r^2 +
 \frac{m^2}{4})}{r^2}\,(X_{,r}^2 - r^{-2}X_{,\theta}^2 + Y_{,r}^2 -
 r^{-2}Y_{,\theta}^2 )\nonumber
\\
		&\qquad\qquad + \frac{2\cos \theta\,(r^2 - \frac{m^2}{4})}{r^3}(X_{,r}\,X_{,\theta} + Y_{,r}\,Y_{,\theta})\Big]
	\;,\label{lambda,r}
\\
 &\mu_{,\theta} = \frac{\rho\,r}{4X^2\Xi}\Big[-\frac{\cos \theta\,(r^2 -
 \frac{m^2}{4})}{r^2}(X_{,r}^2 - r^{-2}X_{,\theta}^2 + Y_{,r}^2 -
 r^{-2}Y_{,\theta}^2 )\nonumber
\\
		&\qquad\qquad + \frac{2\sin \theta(r^2 + \frac{m^2}{4})}{r^3}(X_{,r}\,X_{,\theta} + Y_{,r}\,Y_{,\theta})\Big]
	\;;\label{lambda,theta}
\end{align}
recall that $\Xi$ has been defined in \eq{14VII11.11}.

A consequence of the regularity assumption and \eqref{Xrt}-\eqref{lambda,theta} is that at $\Sigma$, where $r = \frac{m}{2}$, the following identities hold:
\begin{align*}
&X_{,r} = Y_{,r} = \mu_{,r} = \mu_{,\theta}  = 0
	\;,\\
&  X_{,rr} + \frac{2}{m^2}X_{,\theta\theta} + \frac{2\cot\theta}{m^2}X_{,\theta}
	 = \frac{2(X_{,\theta}^2 - Y_{,\theta}^2)}{m^2\,X}
	\;,\\
&  Y_{,rr} + \frac{2}{m^2}Y_{,\theta\theta} + \frac{2\cot\theta}{m^2}Y_{,\theta}
	 = \frac{4X_{,\theta}\,Y_{,\theta}}{m^2\,X}
	\;,\\
&\mu_{,rr}
	= -\frac{1}{m^2\,X^2}(X_{,\theta}^2 + Y_{,\theta}^2)
	\;.
\end{align*}
In particular, $\mu$ is constant along $\Sigma$; we will call its value there $\mu_\Sigma$.

The induced metric on $\Sigma$ is
\[
g_\Sigma = X\,d\varphi^2 + \frac{m^2\sin^2\theta\,e^{2\mu_\Sigma}}{X}\,d\theta^2.
\]
Note that $\partial_\varphi$ vanishes on the axis of rotation. In fact~\cite{ChNguyen}, we can factor $X = \sin^2\theta\,\ringX$ where $\ringX$ is a positive function which is smooth up to $\Sigma$. Then
\begin{equation}
g_\Sigma = \ringX\,\sin^2\theta\,d\varphi^2 + \frac{m^2\,e^{2\mu_\Sigma}}{\ringX}\,d\theta^2
	\label{Eq.IndMetricgSig}
\end{equation}
In order that this metric be smooth at the poles (i.e. $\theta = 0,\pi$), we must have (recall that the integral curves of $\partial_\varphi$ are $2\pi$-periodic)
\bel{14VII11.12}
\ringX|_{\theta = 0} = \ringX|_{\theta = \pi} = m\,e^{\mu_\Sigma}.
\ee
A direct computation leads to the following formula for the scalar curvature $R_\Sigma$ of $g_\Sigma$:
\[
R_\Sigma = \frac{2}{m^2\,e^{2\mu_\Sigma}}\left[\ringX - \frac{1}{2}\ringX_{,\theta\theta} - \frac{3}{2}\cot\theta\,\ringX_{,\theta}\right]
 \;.
\]

The scalar curvature ${}^{(3)}R$ of the slice $\{t = 0\}$ is~\cite{GibbonsHolzegel}
\begin{align*}
{}^{(3)}R
	&= \frac{4X}{e^{2\mu}}\Big[\frac{1}{2}\Delta_\delta(\log\rho - \mu ) - \frac{1}{2}[(\frac{1}{\rho} - \frac{X_{,\rho}}{2X})^2 + \frac{X_{,z}^2}{4X^2}] + \frac{1}{2\rho}(\frac{1}{\rho} + \mu_{,\rho} - \frac{X_{,\rho}}{X})\Big]\\
	&= \frac{2X}{e^{2\mu}}\Big[-(\mu_{,\rho\rho}  + \mu_{,zz}) - [\frac{X_{,\rho}^2}{4X^2} + \frac{X_{,z}^2}{4X^2}]\Big]\\
	&= \frac{2X}{e^{2\mu}}\Big[-\frac{1}{\Xi}(\mu_{,rr}  + r^{-1}\,\mu_{,r} + r^{-2}\mu_{,\theta\theta}) - \frac{1}{4\Xi\,X^2}(X_{,r}^2 + r^{-2}\,X_{,\theta}^2)\Big]
 \;,
\end{align*}
where $\Delta_\delta$ is the Laplace operator of the flat metric on $\R^3$.
At $\Sigma$, we get
\begin{align*}
{}^{(3)}R|_{\Sigma}
	&= \frac{2X}{e^{2\mu_\Sigma}}\Big[-\frac{1}{4\sin^2\theta}\mu_{,rr} - \frac{1}{4m^2\sin^2\theta\,X^2}\,X_{,\theta}^2\Big]\\
	&= \frac{Y_{,\theta}^2}{2m^2\,e^{2\mu_\Sigma}\,\sin^4\theta\,\ringX}.
\end{align*}

We now assume that $\Sigma$ is totally geodesic and weakly stable, under axisymmetric deformations, as a minimal surface in the slice $\{t = 0\}$:
\begin{align*}
0
	&\leq
 \int_\Sigma \left[|\nabla_{g_\Sigma} \phi|^2 + \frac 12 (R_\Sigma - {}^{(3)}R)\,\phi^2\right]\,dv_{g_\Sigma}\\
	&= \frac{2\pi}{m\,e^{\mu_\Sigma}}\int_0^{\pi} \left[\ringX\,\phi_{,\theta}^2
		+ \Big(\left[\ringX - \frac{1}{2}\ringX_{,\theta\theta} - \frac{3}{2}\cot\theta\,\ringX_{,\theta}\right] - \frac{Y_{,\theta}^2}{4\,\sin^4\theta\,\ringX}\Big)\phi^2\right]\,
\sin\theta\,d\theta
\end{align*}
for all $\phi \in C^\infty(\Sigma), \partial_\varphi \phi = 0$. Following~\cite{DainReiris}, taking $\phi = \ringX^{-1/2}$ in particular gives
\begin{align*}
0
	&\leq \int_0^{\pi} \left[\frac{\ringX_{,\theta}^2 }{4\ringX^2}
		+ \left(\ringX - \frac{1}{2}\ringX_{,\theta\theta} - \frac{3}{2}\cot\theta\,\ringX_{,\theta}\right)\ringX^{-1} - \frac{Y_{,\theta}^2}{4\,\sin^4\theta\,\ringX^2}\right]\,\sin\theta\,d\theta \\
	&= \int_0^{\pi} \left[-\frac{\ringX_{,\theta}^2 }{4\ringX^2}
		+ \left(1 - \frac{1}{2}(\log\ringX)_{,\theta\theta} - \frac{3}{2}\cot\theta\,(\log\ringX)_{,\theta}\right) - \frac{Y_{,\theta}^2}{4\,\sin^4\theta\,\ringX^2}\right]\,\sin\theta\,d\theta \\
	&= 2 + \left[\log\ringX|_{\theta = 0} + \log \ringX|_{\theta = \pi}\right]
		+ \int_0^{\pi} \left[-\frac{\ringX_{,\theta}^2 }{4\ringX^2}
		- \log\ringX - \frac{Y_{,\theta}^2}{4\,\sin^4\theta\,\ringX^2}\right]\,\sin\theta\,d\theta \\
	&= 2 + 2\log(m\,e^{\mu_\Sigma})
		+ \int_0^{\pi} \left[-\frac{\ringX_{,\theta}^2 }{4\ringX^2}
		- \log\ringX - \frac{Y_{,\theta}^2}{4\,\sin^4\theta\,\ringX^2}\right]\,\sin\theta\,d\theta
 \;,
\end{align*}
where we have used \eq{14VII11.12} in the last step. Recalling \eqref{Eq.IndMetricgSig}, we have thus arrived at
\bel{15VII11.11}
8(1 + \log \frac{A_\Sigma}{4\pi}) \geq \int_0^{\pi} \left[\frac{\ringX_{,\theta}^2 }{\ringX^2}
		+ 4\log\ringX + \frac{Y_{,\theta}^2}{\sin^4\theta\,\ringX^2}\right]\,\sin\theta\,d\theta =: I[\log\ringX,Y]
	\;,
\ee
where $A_\Sigma$ is the area of $\Sigma$ with respect to $g_\Sigma$.

To make contact with  \cite{HennigAnsorgCederbaum}, let a new coordinate $R$ be defined as
$$
 R= \frac 12 \big(r+\frac{m^2}{4 r}\big)
 \;.
$$
The metric \eq{14VII11.11} then takes the form
\[
g = -\frac{\rho^2}{X}\,dt^2 + \underbrace{X}_{=: \hat u\,\sin^2\theta}\,(d\varphi - \omega\,dt)^2 + \underbrace{\frac{e^{2\mu}}{X}\,\Xi\,r^2}_{=:\hat\mu}\,\left(\frac{dR^2}{R^2 - \frac{m^2}{4}} + d\theta^2\right)
\;.
\]
We wish to show that the key inequality needed in  \cite{HennigAnsorgCederbaum}, namely
\[
\int_0^\pi (\hat u\,\hat \mu)_{,R}\Big|_{\mcH}\,\sin\theta\,d\theta > 0
	\;,
\]
is equivalent to \emph{strict} inequality in \eq{15VII11.11}.
We have
\[
\hat u\,\hat \mu = \frac{r^2\,e^{2\mu}\,\Xi}{\sin^2\theta} = \frac{e^{2\mu}\big[\big(r^2\,\cos 2\theta -  \frac{m^2}{4}\big)^2 + r^4\,\sin^2 2\theta\big]}{r^2\,\sin^2\theta}
\]
and
\begin{align*}
(\hat u\,\hat \mu)_{,R}
	&= \frac{2r^2}{(r^2 - \frac{m^2}{4})\sin^2\theta}\,\Big(\frac{e^{2\mu}\big[\big(r^2\,\cos 2\theta -  \frac{m^2}{4}\big)^2 + r^4\,\sin^2 2\theta\big]}{r^2}\Big)_{,r}\\
	&= \frac{4 }{(r^2 - \frac{m^2}{4})\sin^2\theta}\, {e^{2\mu }\,\mu_{,r}\big[\big(r^2\,\cos 2\theta -  \frac{m^2}{4}\big)^2 + r^4\,\sin^2 2\theta\big]} \\
		&\qquad + \frac{4}{\sin^2\theta}\,e^{2\mu}\frac{r^2 + \frac{m^2}{4}}{r}
 \;.
\end{align*}
So, since $r=m/2$ on the horizon,
\[
(\hat u\,\hat \mu)_{,R}\Big|_{\mcH}
	= m^3\,e^{2\mu_\Sigma}\,\mu_{,rr}\big|_{\mcH}
		 + \frac{4m\,e^{2\mu_\Sigma}}{\sin^2\theta}
 \;.
\]
Recalling the equation for $\mu_{,rr}\big|_{\mcH}$ we see that
\begin{eqnarray*}
 \lefteqn{
(\hat u\,\hat \mu)_{,R}\Big|_{\mcH}
	= -m\,e^{2\mu_\Sigma}\,\frac{1}{X^2}(X_{,\theta}^2 + Y_{,\theta}^2)\big|_{\mcH}
		 + \frac{4m\,e^{2\mu_\Sigma}}{\sin^2\theta}}
&&
 \\
	&&= -m\,e^{2\mu_\Sigma}\,\Big(4\cot^2\theta + 4\cot\theta\,(\log \ringX)_{,\theta} + (\log \ringX)_{,\theta}^2 +
 \frac{Y_{,\theta}^2}{\sin^4\theta\,\ringX^2}\Big)\Big|_{\mcH}
\\
 &&
 \phantom{= \, }
		 + \frac{4m\,e^{2\mu_\Sigma}}{\sin^2\theta}
\\
	&&= -m\,e^{2\mu_\Sigma}\,\Big(4\cot\theta\,(\log \ringX)_{,\theta}  + (\log \ringX)_{,\theta}^2 + \frac{Y_{,\theta}^2}{\sin^4\theta\,\ringX^2}\Big)\Big|_{\mcH}
		 + 4m\,e^{2\mu_\Sigma}
\;.
\end{eqnarray*}
Integration by parts leads indeed to \eq{15VII11.11} with a \emph{strict} inequality.
This is all that was required to establish the angular momentum -- area inequality in \cite{HennigAnsorgCederbaum}. However, the MOTS stability argument presented in the body of our work does not necessarily lead to a strict inequality, and whether the analysis in \cite{HennigAnsorgCederbaum} works under equality in \eqref{15VII11.11} is not clear.

On the other hand, it was shown in \cite[Lemma 4.1]{ADG}
that, provided $J_\Sigma \neq 0$,
\begin{equation}
I[\log\ringX,Y]
	\geq 8\left[\log \frac{\left|Y|_{\theta = 0} - Y|_{\theta = \pi}\right|}{4} + 1\right]
	= 8\left(1 + \log (2|J_\Sigma|)\right)
	\;,\label{Eq.DainIneql}	
\end{equation}
where $J_\Sigma$ is the Komar angular momentum associated to $\Sigma$. Combining \eq{15VII11.11} and \eq{Eq.DainIneql} we have therefore rederived the area-angular momentum inequality of \cite{ADG} (compare \cite{DainReiris,JRD}) in our setting:

\begin{proposition}\label{Prop:AJ}
If the metric $g$ given in \eqref{Eq.MetricgLoc} is regular in $\mcMloc$ up to its ``event horizon'' $\mcH$ and if $\Gamma \approx \Sigma$ is totally geodesic and  stable as a minimal surface in $\{t = 0\} \subset \mcMloc$, then the area $A_\Sigma$ and angular momentum $J_\Sigma$ enjoy the inequality
\[
A_\Sigma \geq 8\pi|J_\Sigma|.
\]
\end{proposition}

In the rest of this section we provide for completeness a proof, in our setting, alternative to \cite{ADG}, of   \eqref{Eq.DainIneql}.
A direct variational approach runs into various difficulties, and therefore
we use a  strategy similar to that in~\cite{HennigAnsorgCederbaum}. It suffices to show that, for all pairs $(\mathring X, Y)$ such that $\log  \mathring X,Y\in C^1([0,\pi])$ and $Y(\pi)-Y(0)=8J_\Sigma $ we have the inequality
\begin{equation}
   \lim_{\epsilon\rightarrow 0} I_\epsilon[\log \ringX,Y] \geq 8 + 8\log(2|J_\Sigma|)
	\;,\label{Eq.XYZ}
\end{equation}
where
\[
I_\epsilon[\log \ringX,Y] = \int_\epsilon^{\pi - \epsilon} \left[\frac{\ringX_{,\theta}^2 }{\ringX^2}
		+ 4\log\ringX + \frac{Y_{,\theta}^2}{\sin^4\theta\,\ringX^2}\right]\,\sin\theta\,d\theta
	\;.
\]
(Note that $I [\log \ringX,Y]=   \lim_{\epsilon\rightarrow 0} I_\epsilon[\log \ringX,Y] $ trivially in the class of functions under consideration.)
We compute
\begin{align*}
I_\epsilon[\log \ringX,Y]
	&= 4\cos\epsilon[\log\ringX(\epsilon) + \log\ringX(\pi - \epsilon)] \\
		&\qquad\qquad + \int_\epsilon^{\pi - \epsilon} \left[\frac{\ringX_{,\theta}^2 }{\ringX^2} + 4(\log\ringX)_{,\theta}\,\cot\theta + \frac{Y_{,\theta}^2}{\sin^4\theta\,\ringX^2}\right]\,\sin\theta\,d\theta\\
	&= 4\cos\epsilon[\log\ringX(\epsilon) + \log\ringX(\pi - \epsilon)]
		+ 4\Big(2\cos\epsilon + \log\frac{1-\cos\epsilon}{1 + \cos\epsilon}\Big)	 \\
		&\qquad\qquad + \int_\epsilon^{\pi - \epsilon} \frac{X_{,\theta}^2  + Y_{,\theta}^2}{X^2}\,\sin\theta\,d\theta\\
	&= 8\log(m\,e^{\mu_\Sigma}) + 8 + 4\,\log\frac{\sin^2\epsilon}{4} + o(1)\\
		&\qquad\qquad + \int_\epsilon^{\pi - \epsilon} \frac{X_{,\theta}^2  + Y_{,\theta}^2}{X^2}\,\sin\theta\,d\theta,
\end{align*}
where in the last identity we have used \eqref{14VII11.12} and the ``little $o$'' notation is understood for small $\epsilon$. To proceed we perform a change of variable $\tau = \log \frac{1 - \cos\theta}{1 + \cos\theta}$ in the integral on the right-hand side:

\begin{align*}
\int_\epsilon^{\pi - \epsilon} \frac{X_{,\theta}^2  + Y_{,\theta}^2}{X^2}\,\sin\theta\,d\theta
	&= 2\int_{-|\tau(\epsilon)|}^{|\tau(\epsilon)|} \frac{X_{,\tau}^2  + Y_{,\tau}^2}{X^2}\,d\tau\\
	&\geq \frac{1}{|\tau(\epsilon)|}\left\{\int_{-|\tau(\epsilon)|}^{|\tau(\epsilon)|} \frac{\sqrt{X_{,\tau}^2  + Y_{,\tau}^2}}{X}\,d\tau\right\}^2.
\end{align*}
It is readily seen that the integral on the right-hand side is no larger than the geodesic distance in the hyperbolic plane $\HH = \{a + i\,b: a > 0, b \in \RR\}$ between $X(\epsilon) + iY(\epsilon)$ and $X(\pi - \epsilon) + iY(\pi - \epsilon)$~\cite[Chapter I]{Katok}:

\begin{align*}
\int_{-|\tau(\epsilon)|}^{|\tau(\epsilon)|} \frac{\sqrt{X_{,\tau}^2  + Y_{,\tau}^2}}{X}\,d\tau
	&\geq {\rm arccosh}\Big(1 + \frac{\sin^4\epsilon[\ringX(\epsilon) - \ringX(\pi - \epsilon)]^2 + [Y(\epsilon) - Y(\pi - \epsilon)]^2}{2\sin^4\epsilon\,\ringX(\epsilon)\,\ringX(\pi - \epsilon)}\Big)\\
	&= \log\frac{64|J_\Sigma|^2}{\sin^4\epsilon\,m^2\,e^{2\mu_\Sigma}} + o(1).
\end{align*}
This implies that
\begin{align*}
\int_\epsilon^{\pi - \epsilon} \frac{X_{,\theta}^2  + Y_{,\theta}^2}{X^2}\,\sin\theta\,d\theta
	 \geq \;
 &
  \frac{1}{|\log\frac{\sin^2\epsilon}{4}|}\left\{\log\frac{1}{\sin^4\epsilon} + \log \frac{64|J_\Sigma|^2}{m^2\,e^{2\mu_\Sigma}}\right\}^2 + o(1)
\\
	 = \;
  &
  \frac{1}{\log\frac{1}{\sin^2\epsilon}}\Big(1 - \frac{\log 4}{\log\frac{1}{\sin^2\epsilon}}\Big)\bigg\{\log^2\frac{1}{\sin^4\epsilon}
\\
  &+ 2\log\frac{1}{\sin^4\epsilon}\,\log \frac{64|J_\Sigma|^2}{m^2\,e^{2\mu_\Sigma}}\bigg\} + o(1)
  \\
	=  \;
 &
  2\log\frac{1}{\sin^4\epsilon} + 4\log\frac{16|J_\Sigma|^2}{m^2\,e^{2\mu_\Sigma}} + o(1).
\end{align*}
Altogether, we obtain
\begin{align*}
I_\epsilon[\log \ringX,Y]
	&\geq 8\log(m\,e^{\mu_\Sigma}) + 8 + 4\log\frac{\sin^2\epsilon}{4} 	
		+ 2\log\frac{1}{\sin^4\epsilon} + 4\log\frac{16|J_\Sigma|^2}{m^2\,e^{2\mu_\Sigma}} + o(1)\\
	&= 8 +  8\log(2|J_\Sigma|) + o(1).
\end{align*}
This establishes \eqref{Eq.XYZ}, whence Proposition \ref{Prop:AJ}.

\bigskip

%----------------------------------------------------------------------------%

\section{Construction of a spacelike hypersurface}
\label{AppSurfConstr}

We would like to show the existence of a spacelike hypersurface $\hyp$ in the d.o.c.\ of a two-Kerr spacetime such that $\hyp$ has compact boundary on a degenerate component of $\partial J^-(\Mext) \cap I^+ (\Mext)$  and coincides with the surface $\{t=0\}$ near the non-degenerate component of the event horizon, if the case occurs.

Recall that the metric takes the form
\[
g= f^{-1}( h (d\rho^2 + dz^2 ) + \rho^2 d\varphi^2 ) - f (dt+ a d\varphi)^2
	\;.
\]
We assume that the degenerate horizon $\mcHdeg$ is located at $\rho = z = 0$ and the non-degenerate horizon is located outside of $\{\rho^2 + z^2 \leq d^2\}$ for some $d > 0$.

Define $(r,\theta)$ by $(\rho,z) = (r\sin\theta, r\cos\theta)$ and let $m_0$ be the Komar mass of $\mcHdeg$. By H\'aj\'i\v{c}ek's theorem on the near horizon geometry at a regular degenerate horizon \cite{Hajicek3Remarks} (see also \cite{LP1}), the angular velocity of $\mcHdeg$ is equal to that of an extreme Kerr with the same mass, i.e. $a = - 2m_0$ on $\mcHdeg$. Performing a change of variable $\phi = \varphi - \frac{1}{2m_0}\,t$ and $u = t$, we obtain
\begin{align*}
g
	&= f^{-1}\, h (dr^2 + r^2\,d\theta^2) + (\rho^2\,f^{-1} - f\,a^2) d\phi^2\\
		&\qquad\qquad + 2\Big[\frac{1}{2m_0}(\rho^2\,f^{-1} - f\,a^2) - f\,a\Big]\,du\,d\phi\\
		&\qquad\qquad - \Big[-\frac{1}{4m_0^2}(\rho^2\,f^{-1} - f\,a^2) + \frac{1}{m_0}\,f\,a + f\Big]\,du^2\\
	&= f^{-1}\, h (dr^2 + r^2\,d\theta^2) + (\rho^2\,f^{-1} - f\,a^2) \Big\{d\phi + \Big[\frac{1}{2m_0} - \frac{f\,a}{\rho^2\,f^{-1} - f\,a^2}\Big]\,du\Big\}^2\\
		&\qquad\qquad - \frac{\rho^2}{\rho^2\,f^{-1} - f\,a^2}\,du^2
	\;.
\end{align*}
Note that $\partial_u$ is Killing and tangential to $\mcHdeg$. Thus, the near horizon geometry of $g$ at $\mcHdeg$ can be computed by doing a scaling $r \mapsto \lambda\,r$ and $u \mapsto \frac{u}{\lambda}$ and taking the limit $\lambda \rightarrow 0$:
\begin{align*}
g_{NH}
	&= \lim_{\lambda \rightarrow 0}\Big\{ f(\lambda r, \theta)^{-1}\,h(\lambda r, \theta)(\lambda^2\,dr^2 + \lambda^2\,r^2\,d\theta)\\
		&\qquad + (\lambda^2\,r^2\,\sin^2\theta\,f(\lambda r, \theta)^{-1} - f(\lambda r, \theta)\,a(\lambda r, \theta)^2)\\
			&\qquad\qquad \Big\{d\phi + \Big[\frac{1}{2m_0} - \frac{f(\lambda r, \theta)\,a(\lambda r, \theta)}{\lambda^2\,r^2\,\sin^2\theta\,f(\lambda r, \theta)^{-1} - f(\lambda r, \theta)\,a(\lambda r, \theta)^2}\Big]\,du\Big\}^2\\
		&\qquad - \frac{\lambda^2\,r^2\,\sin^2\theta}{\lambda^2\,r^2\,\sin^2\theta\,f(\lambda r, \theta)^{-1} - f(\lambda r, \theta)\,a(\lambda r, \theta)^2}\,du^2\Big\}
	\;.
\end{align*}
On the other hand, again by H\'aj\'i\v{c}ek's theorem, this process always leads to that of the corresponding extreme Kerr, i.e.
\begin{align*}
g_{NH}
	&= \frac{m_0^2}{r^2}(1+ \cos^2\theta)[dr^2 + r^2\,d\theta^2] + \frac{4m_0^2\sin^2\theta}{1 + \cos^2\theta}\,\Big[d\phi + \frac{r}{2m_0^2}\,du\Big]^2\\
		&\qquad\qquad - \frac{r^2}{4m_0^2}\,(1 + \cos^2\theta)\,du^2
	\;.
\end{align*}
In fact, the analysis in \cite{ChNguyen} (see in particular Sections 2 and 3.1) shows that we have the following asymptotics as $r \rightarrow 0$,
\begin{align}
&f^{-1}h = \frac{m_0^2}{r^2}(1 + \cos^2\theta)\,\big[1 + r\,P(r,\theta)\big]
	\;,\label{Eq.Asymp1}\\
&\rho^2\,f^{-1} - f\,a^2 = \frac{4m_0^2\sin^2\theta}{1 + \cos^2\theta}\big[1 + r\,Q(r,\theta)\big]
	\;,\label{Eq.Asymp2}\\
&\frac{1}{2m_0} - \frac{f\,a}{\rho^2\,f^{-1} - f\,a^2} = \frac{r}{2m_0^2}\big[1 + r\,S(r,\theta)\big]
	\;,\label{Eq.Asymp3}
\end{align}
where $P$, $Q$ and $S$ are smooth functions satisfying
\[
1 + r\,P(r,\theta) > 0 \text{ and } 1 + r\,Q(r,\theta) > 0.
\]

We have thus obtained the following representation of $g$:
\begin{align}
g
	&=  \frac{m_0^2}{r^2}(1 + \cos^2\theta)\,\big[1 + r\,P(r,\theta)\big](dr^2 + r^2\,d\theta^2)
	\nonumber\\
		&\qquad\qquad + \frac{4m_0^2\sin^2\theta}{1 + \cos^2\theta}\big[1 + r\,Q(r,\theta)\big] \Big\{d\phi + \frac{r}{2m_0^2}\big[1 + r\,S(r,\theta)\big]\,du\Big\}^2
	\nonumber\\
		&\qquad\qquad - \frac{r^2}{4m_0^2}\,(1 + \cos^2\theta)\,\frac{1}{1 + r\,Q(r,\theta)}\,du^2
	\;.\label{Eq.gExp}
\end{align}

We note that the surface $\{u = 0\}$ is the same as $\{t = 0\}$. The desired hypersurface $\hyp$ will take the form
\[
\hyp = \{(u,\rho,z,\varphi): u = U(r)\}
\]
where $U$ is smooth function for $r > 0$. In fact, we pick $U = U_0\,\chi$ where
\[
U_0 = \frac{2m_0^2}{r} - \frac{4m_0^{3/2}}{r^{1/2}}
	\;,
\]
and $\chi$ is a smooth non-increasing cut-off function such that $\chi(r) = 0$ for $r > \delta$ for some $\delta$ to be specified, $\chi(0) = 1$ and $\chi^{(k)}(0) = 0$ for any $k \geq 1$.

Let us first show that such surface will intersect $\mcHdeg$ along a smooth compact boundary. The induced metric on $\hyp$ is
\begin{align}
{}^{(3)}g
	&=  \frac{m_0^2}{r^2}(1 + \cos^2\theta)\,\big[1 + r\,P(r,\theta)\big](dr^2 + r^2\,d\theta^2)
	\nonumber\\
		&\qquad\qquad + \frac{4m_0^2\sin^2\theta}{1 + \cos^2\theta}\big[1 + r\,Q(r,\theta)\big] \Big\{d\phi + \frac{r}{2m_0^2}\big[1 + r\,S(r,\theta)\big]\,U_{,r}\,dr\Big\}^2
	\nonumber\\
		&\qquad\qquad - \frac{r^2}{4m_0^2}\,(1 + \cos^2\theta)\,\frac{1}{1 + r\,Q(r,\theta)}\,U_{,r}^2\,dr^2
	\;.\label{Eq.3gExp}
\end{align}
Using that $U = U_0\,\chi$, we see that
\begin{align*}
{}^{(3)}g
	&=  \frac{m_0^2}{r^2}(1 + \cos^2\theta)\,\big[1 + O(r)\big](dr^2 + r^2\,d\theta^2)
	\\
		&\qquad\qquad + \frac{4m_0^2\sin^2\theta}{1 + \cos^2\theta}\big[1 + O(r)\big] \Big\{d\phi - \frac{1}{r}\big[1 - \frac{r^{1/2}}{m_0^{1/2}} + O(r)\big]\,dr\Big\}^2
	\\
		&\qquad\qquad - \frac{m_0^2}{r^2}\,(1 + \cos^2\theta)\,\big[1 - \frac{2r^{1/2}}{m_0^{1/2}} + O(r)\big]\,dr^2\\
	&=  m_0^2(1 + \cos^2\theta)\,\big[(\frac{2r^{-3/2}}{m_0^{1/2}} + O(r^{-1}))dr^2 + d\theta^2\big]
	\\
		&\qquad\qquad + \frac{4m_0^2\sin^2\theta}{1 + \cos^2\theta}\big[1 + O(r)\big] \Big\{d\phi - \frac{1}{r}\big[1 - \frac{r^{1/2}}{m_0^{1/2}} + O(r)\big]\,dr\Big\}^2
	\;.
\end{align*}
It is readily seen that every curve of the form $\{\phi = \log r, 0 < r \leq \delta\}$ has finite length, whence $\{r = 0\}$ corresponds to a boundary of $\hyp$.

It remains to show that $\hyp$ is spacelike. This will depend on certain property of the cut-off function $\chi$. First, choose $\delta > 0$ sufficiently small such that
\begin{itemize}
  \item $U_0(r) > 0$ for $0 < r \leq 5\delta$,
  \item $U_0'(r) < 0$ for $0 < r \leq 5\delta$,
  \item $-\frac{2m_0^2}{r^2}\,\big[1 + r\,P(r,\theta)\big]^{1/2}\big[1 + r\,Q(r,\theta)]^{1/2} < -\frac{2m_0^2}{r^2} + \frac{m_0^{3/2}}{r^{3/2}}$ for $0 \leq r \leq 5\delta$.
\end{itemize}
We claim that $\chi$ can be selected such that $\chi$ is decreasing and
\begin{equation}
U' = (U_0\,\chi)' > -\frac{2m_0^2}{r^2} + \frac{m_0^{3/2}}{r^{3/2}} \text{ for } 0 < r < 5\delta\;.
	\label{Eq.KeyCutoffProp}
\end{equation}

To see this, consider first a Lipschitz cut-off function $\bar\chi$ defined by
\[
\bar\chi(r) = \left\{\begin{array}{l}
	\frac{\delta - r}{\delta} \text{ for } 0 \leq r \leq \delta
	\;,\\
	0 \text{ for } r > \delta
	\;.
\end{array}\right.
\]
Then, for $0 < r < \delta$,
\begin{align*}
U_0'\,\bar\chi + U_0\,\bar\chi'
	&= \Big(-\frac{2m_0^2}{r^2} + \frac{2m_0^{3/2}}{r^{3/2}}\Big)\frac{\delta - r}{\delta} - \Big(\frac{2m_0^2}{r} - \frac{4m_0^{3/2}}{r^{1/2}}\Big)\,\frac{1}{\delta}\\
	&= -\frac{2m_0^2}{r^2} + \frac{2m_0^{3/2}}{r^{3/2}} + \frac{2m_0^{3/2}}{r^{1/2}}\,\frac{1}{\delta}\\
	&> -\frac{2m_0^2}{r^2} + \frac{2m_0^{3/2}}{r^{3/2}}
	\;.
\end{align*}

We would like next to smoothen $\bar\chi$ to obtain a desired $\chi$. We start with a $C^{1,1}$ smoothing; once this is done, the rest is routine. For some $\mu, \mu' \in (0,\frac{1}{2})$ to be specified, define $\tilde\chi$ by
\[
\tilde\chi(r) = \left\{\begin{array}{l}
	1 \text{ for } r = 0
	\;,\\
	1 - \mu\,\exp\big(1 - \frac{\mu\delta}{r}\big) \text{ for } 0 < r \leq \mu\delta
	\;,\\
	\bar\chi(r) \text{ for } \mu\delta \leq r \leq (1 - \mu')\delta
	\;,\\
	\frac{1}{4\mu'\,\delta^2}(r - (1 + \mu')\delta)^2 \text{ for } (1 - \mu')\delta \leq r \leq (1 + \mu')\delta
	\;,\\
	0 \text{ for } r \geq (1 + \mu')\delta
	\;.
\end{array}\right.
\]
A direct computation shows that $\tilde\chi$ is $C^{1,1}$.

For $0 < r \leq \mu\delta$, we have
\begin{align*}
U_0'\,\tilde\chi + U_0\,\tilde\chi'
	&= -\frac{2m_0^2}{r^2} + \frac{2m_0^{3/2}}{r^{3/2}}\Big[1 - \mu\,\exp\Big(1 - \frac{\mu\delta}{r}\Big)\Big]\\
		&\qquad - \mu\,\exp\Big(1 - \frac{\mu\delta}{r}\Big) \Big[-\frac{2m_0^2}{r^2} + \Big(\frac{2m_0^2}{r} - \frac{4m_0^{3/2}}{r^{1/2}}\Big)\,\frac{\mu\delta}{r^2}\Big]\\
	&> -\frac{2m_0^2}{r^2} + \frac{2m_0^{3/2}}{r^{3/2}}\big[1 - \mu\big]\\
		&\qquad + \frac{2m_0^2\mu^{1/2}}{\delta^{1/2}r^{3/2}}\,\frac{\mu^{1/2}\,\delta^{1/2}}{r^{1/2}}\,\Big(1 - \frac{\mu\delta}{r}\Big)\exp\Big(1 - \frac{\mu\delta}{r}\Big) \\
	&\geq -\frac{2m_0^2}{r^2} + \frac{2m_0^{3/2}}{r^{3/2}}\big[1 - \mu - \frac{m_0^{1/2}\mu^{1/2}\,K}{\delta^{1/2}}\big]
	\;,
\end{align*}
where
\[
K = \sup_{0 < x \leq 1} x^{-1/2}\,(-1 + x^{-1})\,\exp(1 - x^{-1}) < \infty
	\;.
\]
Thus, if we choose $\mu \ll \min(\delta,\frac{1}{20})$ sufficiently small, then
\[
U_0'\,\tilde\chi + U_0\,\tilde\chi' > -\frac{2m_0^2}{r^2} + \frac{3m_0^{3/2}}{2r^{3/2}} \text{ for } 0 < r \leq \mu\delta
	\;.
\]

For $(1 - \mu')\,\delta \leq r \leq (1 + \mu')\delta$, we have
\begin{align*}
U_0'\,\tilde\chi + U_0\,\tilde\chi'
	&= \tilde\chi\Big(-\frac{2m_0^2}{r^2} + \frac{2m_0^{3/2}}{r^{3/2}}\Big) + r\tilde\chi'\Big(\frac{2m_0^2}{r^2} - \frac{4m_0^{3/2}}{r^{3/2}}\Big)\\
	&= -(\tilde\chi - r\,\tilde\chi')\,\frac{2m_0^2}{r^2} + (\tilde\chi - 2r\tilde\chi')\frac{2m_0^{3/2}}{r^{3/2}}\\
	&= -\frac{2m_0^2}{r^2} + \Big[\frac{m_0^{1/2}}{r^{1/2}} + (-\frac{m_0^{1/2}}{r^{1/2}} + 1)\tilde\chi - r(-\frac{m_0^{1/2}}{r^{1/2}} + 2)\tilde\chi'\Big]\frac{2m_0^{3/2}}{r^{3/2}}
%	&> - \frac{2m_0^2}{r^2} + \frac{3m_0^{3/2}}{2r^{3/2}}
	\;.
\end{align*}
The term in the square bracket is estimated as follows:
\begin{align*}
&\frac{m_0^{1/2}}{r^{1/2}} + (-\frac{m_0^{1/2}}{r^{1/2}} + 1)\tilde\chi - r(-\frac{m_0^{1/2}}{r^{1/2}} + 2)\tilde\chi' \\
	&\qquad =\frac{m_0^{1/2}}{r^{1/2}} + \frac{r - (1 + \mu')\delta}{4\mu'\,\delta^2}\Big[(1+\mu')m_0^{1/2}\delta\,r^{-1/2} - (1+\mu')\delta + m_0^{1/2}r^{1/2} - 3r  \Big]\\
	&\qquad \geq \frac{m_0^{1/2}}{(1+\mu')^{1/2}\,\delta^{1/2}} - \frac{1}{2\delta}\Big[(1+\mu')(1-\mu')^{-1/2}m_0^{1/2}\delta^{1/2} - (1+\mu')\delta \\
		&\qquad\qquad\qquad + (1+\mu')^{1/2}\,m_0^{1/2}\,\delta^{1/2} - 3(1-\mu')\,\delta  \Big]\\
	&\qquad = \frac{m_0^{1/2}}{\delta^{1/2}}\Big[\frac{1}{(1+\mu')^{1/2}} - \frac{1}{2}(1+\mu')(1-\mu')^{-1/2} - \frac{1}{2}(1+\mu')^{1/2}\Big] + 2 - \mu'\\
	&\qquad \geq \frac{-2m_0^{1/2}\mu'}{\delta^{1/2}} + 2 - \mu'\\
	&\qquad > \frac{3}{2} \text{ for all sufficiently small $\mu' \ll \delta^{1/2} \ll 1$},
\end{align*}
where in the second-to-last inequality we have used the expansion
\[
\frac{1}{(1+x)^{1/2}} - \frac{1}{2}(1+x)(1-x)^{-1/2} - \frac{1}{2}(1+x)^{1/2} = - \frac{3}{2}x + o(x) \text{ as } x\rightarrow 0.
\]
Altogether, we have shown that
\[
U_0'\,\tilde\chi + U_0\,\tilde\chi' > -\frac{2m_0^2}{r^2} + \frac{3m_0^{3/2}}{2r^{3/2}} \text{ for } 0 < r < 5\delta
	\;.
\]
Since $\tilde\chi$ is $C^{1,1}$, a routine smoothing procedure then gives a cut-off function $\chi$ which satisfies \eqref{Eq.KeyCutoffProp} as claimed.

From our choice of $\chi$, we see that
\[
0 > U_{,r} >  -\frac{2m_0^2}{r^2}\,\big[1 + r\,P(r,\theta)\big]^{1/2}\big[1 + r\,Q(r,\theta)]^{1/2}
\]
for $0 \leq r \leq 5\delta$. This implies that
\[
\frac{m_0^2}{r^2}(1 + \cos^2\theta)\,\big[1 + r\,P(r,\theta)\big] - \frac{r^2}{4m_0^2}\,(1 + \cos^2\theta)\,\frac{1}{1 + r\,Q(r,\theta)}\,U_{,r}^2 > 0
\]
for any $0 < r \leq d$. Lowering $\delta$ again if necessary, it is then easy to check that the metric ${}^{(3)}g$ given in \eqref{Eq.3gExp} is Riemannian. This finishes the construction of $\hyp$ which was needed in Section \ref{S28VII11.1} in the case where exactly one component of the horizon is degenerate.

\bigskip

\noindent{\sc Acknowledgements.} We acknowledge many useful comments, email exchanges, and discussions with Michael Eichmair and Joerg Hennig. Special thanks are due to Joerg Hennig for making his {\sc Maple} files available to us.  PTC was supported in part
by the Polish Ministry of Science and Higher Education grant Nr
N N201 372736. ME was partially supported by the Polish Ministry of Science and Higher Education under the grant for young scientists and PhD students. ME and SJS were supported by the John Templeton Foundation. ME, LN and SJS would like to thank the University of Vienna for hospitality and support.

\bibliographystyle{amsplain}
\bibliography{%
../references/reffile,%
../references/newbiblio,%
../references/newbiblio2,%
../references/bibl,%
../references/howard,%
../references/bartnik,%
../references/myGR,%
../references/newbib,%
../references/Energy,%
../references/netbiblio,%
../references/PDE}

\def\polhk#1{\setbox0=\hbox{#1}{\ooalign{\hidewidth
  \lower1.5ex\hbox{`}\hidewidth\crcr\unhbox0}}} \def\cprime{$'$}
  \def\cprime{$'$}
\providecommand{\bysame}{\leavevmode\hbox to3em{\hrulefill}\thinspace}
\providecommand{\MR}{\relax\ifhmode\unskip\space\fi MR }
\providecommand{\MRhref}[2]{%
  \href{http://www.ams.org/mathscinet-getitem?mr=#1}{#2}
}
\providecommand{\href}[2]{#2}
\begin{thebibliography}{10}

\bibitem{ADG}
A.~Acena, S.~Dain, and M.E.~Gabach Cl{\'e}ment, \emph{{Horizon area--angular
  momentum inequality for a class of axially symmetric black holes}}, Class.\
  Quantum Grav. \textbf{28} (2010), 105014, 18, arXiv:1012.2413 [gr-qc].
  \MR{2801106}

\bibitem{AEM}
L.~Andersson, M.~Eichmair, and J.~Metzger, \emph{{Jang's equation and its
  applications to marginally trapped surfaces}},  (2010), arXiv:1006.4601
  [gr-qc].

\bibitem{AMS2}
L.~Andersson, M.~Mars, and W.~Simon, \emph{{Stability of marginally outer
  trapped surfaces and existence of marginally outer trapped tubes}}, Adv.
  Theor. Math. Phys. \textbf{12} (2008), 853--888. \MR{MR2420905 (2010a:83005)}

\bibitem{AndM2}
L.~Andersson and J.~Metzger, \emph{{The area of horizons and the trapped
  region}}, Commun.\ Math.\ Phys. \textbf{290} (2009), 941--972,
  arXiv:0708.4252 [gr-qc]. \MR{MR2525646 (2010f:53118)}

\bibitem{Bardeen}
J.M. Bardeen, \emph{Rapidly rotating stars, disks and black holes}, Black
  holes, Les Houches (C.\ and B.~deWitt, eds.), Gordon Breach, New York, 1973,
  pp.~241�--289.

\bibitem{ChUone}
P.T. Chru\'{s}ciel, \emph{Mass and angular-momentum inequalities for
  axi-symmetric initial data sets. {I. Positivity of mass}}, Annals Phys.
  \textbf{323} (2008), 2566--2590, doi:10.1016/j.aop.2007.12.010,
  arXiv:0710.3680 [gr-qc].

\bibitem{ChCo}
P.T. Chru\'{s}ciel and J.~Lopes Costa, \emph{On uniqueness of stationary black
  holes}, Ast\'erisque \textbf{321} (2008), 195--265, arXiv:0806.0016v2
  [gr-qc].

\bibitem{CGS}
P.T. Chru\'{s}ciel, G.~Galloway, and D.~Solis, \emph{Topological censorship for
  {K}aluza-{K}lein space-times}, Ann.\ Henri Poincar\'e \textbf{10} (2009),
  893--912, arXiv:0808.3233 [gr-qc]. \MR{MR2533875}

\bibitem{ChNguyen}
P.T. Chru\'{s}ciel and L.~Nguyen, \emph{{A uniqueness theorem for degenerate
  Kerr-Newman black holes}}, Ann.\ H.~Poincar\'e \textbf{11} (2010), 585--609,
  arXiv:1002.1737 [gr-qc]. \MR{2677740 (2011i:53123)}

\bibitem{ChNguyenMass}
\bysame, \emph{{A lower bound for the mass of axisymmetric connected black hole
  data sets}}, Class.\ Quantum Grav. \textbf{28} (2011), 125001,
  arXiv:1102.1175 [gr-qc].

\bibitem{DainReiris}
S.~Dain and M.~Reiris, \emph{{Area - Angular momentum inequality for
  axisymmetric black holes}}, Phys.\ Rev.\ Lett. \textbf{107} (2011), 051101,
  arXiv:1102.5215 [gr-qc].

\bibitem{Eichmair}
M.~Eichmair, \emph{The {P}lateau problem for marginally outer trapped
  surfaces}, Jour.\ Diff.\ Geom. \textbf{83} (2009), 551--583, arXiv:0711.4139
  [math.DG]. \MR{2581357 (2011c:53066)}

\bibitem{GallowaySchoen}
G.J. Galloway and R.~Schoen, \emph{A generalization of {Hawking's} black hole
  topology theorem to higher dimensions}, Commun.\ Math.\ Phys. \textbf{266}
  (2006), 571--576, arXiv:gr-qc/0509107. \MR{2238889 (2007i:53078)}

\bibitem{GibbonsHolzegel}
G.W. Gibbons and G.~Holzegel, \emph{The positive mass and isoperimetric
  inequalities for axisymmetric black holes in four and five dimensions},
  Class.\ Quantum Grav. \textbf{23} (2006), 6459--6478, arXiv:gr-qc/0606116.
  \MR{MR2272015}

\bibitem{Hajicek3Remarks}
P.~H{\'a}j{\'{\i}}{\v{c}}ek, \emph{Three remarks on axisymmetric stationary
  horizons}, Commun.\ Math.\ Phys. \textbf{36} (1974), 305--320. \MR{MR0418816
  (54 \#6852)}

\bibitem{HennigAnsorgCederbaum}
J.~Hennig, M.~Ansorg, and C.~Cederbaum, \emph{A universal inequality between
  the angular momentum and horizon area for axisymmetric and stationary black
  holes with surrounding matter}, Class.\ Quantum Grav. \textbf{25} (2008),
  162002, 8. \MR{MR2429717 (2009e:83092)}

\bibitem{HennigNeugebauer2}
J.~Hennig and G.~Neugebauer, \emph{{Non-existence of stationary two-black-hole
  configurations: The degenerate case}}, General Relativity and Gravitation
  \textbf{43} (2011), 3139--3162, arXiv:1103.5248 [gr-qc].

\bibitem{Herzlich:bh}
M.~Herzlich, \emph{The positive mass theorem for black holes revisited}, Jour.\
  Geom.\ Phys. \textbf{26} (1998), 97--111. \MR{1626060 (99g:83029)}

\bibitem{Heusler:book}
M.~Heusler, \emph{Black hole uniqueness theorems}, Cambridge University Press,
  Cambridge, 1996.

\bibitem{JRD}
J.L. Jaramillo, M.~Reiris, and S.~Dain, \emph{{Black hole Area-Angular momentum
  inequality in non-vacuum spacetimes}},  (2011), arXiv:1106.3743 [gr-qc].

\bibitem{Katok}
S.~Katok, \emph{Fuchsian groups}, Chicago Lectures in Mathematics, University
  of Chicago Press, Chicago, IL, 1992. \MR{1177168 (93d:20088)}

\bibitem{Kramer}
D.~Kramer, \emph{Two {K}err-{NUT} constituents in equilibrium}, Gen. Rel.\
  Grav. \textbf{18} (1986), 497--509. \MR{839716 (87g:83031)}

\bibitem{LP1}
J.~Lewandowski and T.~{Paw\l owski}, \emph{Extremal isolated horizons: A local
  uniqueness theorem}, Class.\ Quantum Grav. \textbf{20} (2003), 587--606,
  arXiv:gr-qc/0208032.

\bibitem{HennigNeugebauer}
G.~Neugebauer and J.~Hennig, \emph{Non-existence of stationary two-black-hole
  configurations}, Gen. Relativity Gravitation \textbf{41} (2009), no.~9,
  2113--2130. \MR{MR2534657}

\bibitem{HennigNeugebauer3}
\bysame, \emph{{Stationary two-black-hole configurations: A non-existence
  proof}}, Jour.\ Geom.\ Phys. (2011), in press, arXiv:1105.5830 [gr-qc].

\bibitem{Neugebauer:2003qe}
G.~Neugebauer and R.~Meinel, \emph{Progress in relativistic gravitational
  theory using the inverse scattering method}, Jour.\ Math.\ Phys. \textbf{44}
  (2003), 3407--3429, arXiv:gr-qc/0304086.

\bibitem{RaczWald2}
I.~R{\'a}cz and R.M. Wald, \emph{Global extensions of space-times describing
  asymptotic final states of black holes}, Class.\ Quantum Grav. \textbf{13}
  (1996), 539--552, arXiv:gr-qc/9507055. \MR{MR1385315 (97a:83071)}

\bibitem{Varzugin1}
G.G. Varzugin, \emph{Equilibrium configuration of black holes and the method of
  the inverse scattering problem}, Teoret. Mat. Fiz. \textbf{111} (1997),
  no.~3, 345--355. \MR{1472213 (98e:83087)}

\end{thebibliography}

\end{document}